\documentclass[epj]{svjour}

\usepackage{graphics}

\begin{document}

\title{Performances  of a large mass ZnMoO$_4$ scintillating bolometer for a next generation $0\nu$DBD experiment\\}
\author{
J.~W.~Beeman\inst{1}, 
F.~Bellini\inst{2,3},
C.~Brofferio\inst{4,5},
L.~Cardani\inst{2,3},
N.~Casali\inst{6,7},
O.~Cremonesi\inst{5},
I.~Dafinei\inst{3},
S.~Di~Domizio\inst{8},
F.~Ferroni\inst{2,3},
E.~Gorello\inst{2,3},
E.~N.~Galashov\inst{9},
L.~Gironi\inst{4,5},
S.S.~Nagorny\inst{10},
F.~Orio\inst{3},
M.~Pavan\inst{4,5},
L.~Pattavina\inst{5}
G.~Pessina\inst{5},
G.~Piperno\inst{2,3},
S.~Pirro\inst{5}\thanks{\emph{Corresponding Author}},
E.~Previtali\inst{5},
C.~Rusconi\inst{5},
V.~N.~Shlegel\inst{9},
C.~Tomei\inst{3},
M.~Vignati\inst{3}
}

\authorrunning{J.~W.~Beeman \emph{et al}.}
\titlerunning{Performances  of a large mass ZnMoO$_4$ scintillating bolometer}
\institute{
Lawrence Berkeley National Laboratory, Berkeley, California 94720, USA  					\and 
Dipartimento di Fisica, Sapienza Universit\`{a} di Roma,  I 00185 Roma, Italy 				\and 
INFN, Sezione di Roma, I 00185 Roma, Italy 		   		  		  							\and 
Dipartimento di Fisica, Universit\`{a} di Milano-Bicocca, I 20126 Milano, Italy 			\and 
INFN, Sezione di Milano Bicocca, I 20126 Milano, Italy	  		  		  					\and 
Dipartimento di Fisica, Universit\`{a} degli studi dell'Aquila,  I 67100  L'Aquila, Italy 	\and 
INFN, Laboratori Nazionali del Gran Sasso, I 67010 L'Aquila, Italy 		  					\and 
INFN,  Sezione di Genova,   I 16146 Genova,  Italy 			 								\and 
Nikolaev Institute of Inorganic Chemistry - SB RAS, 630090 Novosibirsk - Russia				\and 
Institute for Nuclear Research 03680 Kyiv, Ukraine   			   	 						 
}

%
%
\abstract{
We present the performances of a 330 g zinc molybdate (ZnMoO$_4$)  crystal working as scintillating bolometer 
as a possible candidate for  a next generation experiment to search for  neutrinoless double beta decay of $^{100}$Mo.  
The energy resolution, evaluated at the 2615 keV $\gamma$-line of $^{208}$Tl, is 6.3 keV FWHM.
The internal radioactive contaminations of the  ZnMoO$_4$  were evaluated  as $<$~6 $\mu$Bq/kg ($^{228}$Th) and 27$\pm$6 $\mu$Bq/kg ($^{226}$Ra). 
We also present the results of the  $\alpha$ vs $\beta/\gamma$  discrimination, obtained through the 
scintillation light as well as through the study of the shape of the thermal signal alone.
\PACS{
      {23.40.Bw}{Weak interactions in $\beta$ decay}   \and
      {29.40.Mc}{Scintillation detectors} \and 
	  {07.57.Kp}{Bolometers} 
	     } 
} 

\maketitle
\section{Introduction}
\label{sec:introduction}
The neutrinoless Double Beta Decay ($0\nu$DBD) is a nuclear process that, if observed, would establish that 
the total lepton number is not a conserved quantity and that the neutrino is a Majorana particle, and would set the 
absolute mass scale of neutrinos~\cite{Vissani,Avi08,Wern-2011}.
Plenty of experiments  are now in the construction phase and many others are in R\&D phase~\cite{Bara-2011}.
Very recently EXO~\cite{EXO-2012} and KamLAND-Zen~\cite{KamZen-2012} set very competitive limits on the $0\nu$DBD  
half-life of $^{136}$Xe.

The main challenges for all the different experimental techniques are the 
same~\cite{EPJA-2006}: i) increase of the active mass, ii) decrease of the background, and iii) improvement of the energy resolution.

Thermal bolometers are  ideal detectors for this kind of research: crystals  can be grown with a variety of  interesting DBD-emitters 
and multi-kg detectors can be operated with excellent energy resolution~\cite{massive-tellurium} which, perhaps, represents one of  
the most critical aspects for next generation experiments.

The Cuoricino experiment~\cite{Andreotti:2010vj} searched for the $0\nu$DBD of  $^{130}$Te operating 62 TeO$_2$ bolometers. 
The  Cuoricino data demonstrated that the background in the region of interest is dominated by radioactive 
contaminations on the surfaces facing the detectors. $\alpha$  particles produced by these contaminants can lose a 
fraction of their energy in the host material, and the rest in the detector, thus producing a flat background from 
the energy of the decay (several MeV) down to the $0\nu$DBD region~\cite{Patta-2011}. 
Moreover  simulations show that this contribution will largely dominate the expected background of  the CUORE 
experiment~\cite{ACryo,Pavan:2008zz} in the region of interest, 
since TeO$_2$ bolometers do not allow to distinguish $\alpha$ particles from the electrons emitted in the $0\nu$DBD.

The natural way to discriminate this background, is to use  scintillating bolometers~\cite{Pirro:2005ar}. 
In such devices the simultaneous and independent readout of the heat and  the scintillation light signals permits to discriminate events 
due to $\beta/\gamma$, $\alpha$ and neutrons interactions  thanks to  their different scintillation yield.

$^{100}$Mo is a very interesting $\beta\beta$-isotope because of its large transition energy $Q_{\beta\beta}=3034 $~keV
and a considerable natural isotopic abundance $\delta=9.67\%$.
Several inorganic scintillators containing molybdenum were developed in the last years. Among them, ZnMoO$_4$ was recently grown~\cite{Ivle08}  
and the first  cryogenic detector gave very promising results~\cite{Gironi:2010hs}. Some scintillating 
crystals,  the molybdates in particular, show a very peculiar feature: the thermal pulse induced by an $\alpha$ particle shows a 
slightly faster decay time with respect to the one induced by  $\gamma$ interactions~\cite{Arnaboldi:2011gj}.
This feature  seems to  be explained~\cite{Gironi:2012js} by the relatively long scintillation decay time  (of the order of 
hundreds of $\mu$s) observed in  some scintillating crystals. This long decay, combined with a high percentage 
of non-radiative  de-excitation of  the scintillation channel, will transfer phonons (i.e. heat) to the crystal.
This extremely tiny, but measurable, time dependent phonon release has a different absolute value for  isoenergetic $\alpha$ and 
$\beta/\gamma$ particles due to their different scintillation yield.

It was very recently measured that the $\alpha$ vs $\gamma$ separation on a 29 g  ZnMoO$_4$ crystal can 
reach an extremely high efficiency~\cite{ZMO-Lucifer} using the pulse shape discrimination (PSD) alone, while the separation based 
on the scintillation light, even on smaller crystals (5 g), shows a smaller efficiency~\cite{Giuliani-2012}.

A next generation experiment, nonetheless, will need to run detectors with a considerably larger mass (of the order of few hundreds of grams each).
It is not  straightforward to foresee the performance of the particle discrimination method over a significantly larger sample.
This is due to the fact that the PSD is sensitive to the signal to noise ratio, and a larger mass of the absorber crystal
leads to a smaller signal amplitude ($\propto$~(detector mass)$^{-1}$).
The same holds (in principle) for the discrimination based on the scintillation light: usually (especially in the case 
of non transparent crystals) the larger the crystal, the smaller the light output. Moreover one has to consider that 
this compound is characterized by an extremely tiny Light Yield (LY): 1$\div$2 keV/MeV.

The purpose of this work is to study the most important parameters (energy resolution, $\alpha$ vs $\gamma$ discrimination, internal 
radiopurity) on a crystal whose size (330 g)  matches very closely the requirement of a next generation $0\nu$DBD experiment.
 
\section{Experimental set-up}

\label{Set-Up}
The 330 g ZnMoO$_4$ crystal   studied in  this work was  grown in the Nikolaev Institute of Inorganic Chemistry (NIIC, Novosibirsk, Russia).
Starting material for the crystal gro\-wth were high purity ZnO (produced by Umicore)  and
MoO$_3$,  synthesized  by NIIC.

Crystals are grown by the low-thermal-gradient Czo\-chral\-ski technique (LTG Cz)~\cite{Galashov-2010}
from a melt contained in a  80~mm diameter  platinum crucible. 
\begin{figure}
\centering
\resizebox{0.35\textwidth}{!}{%
\includegraphics{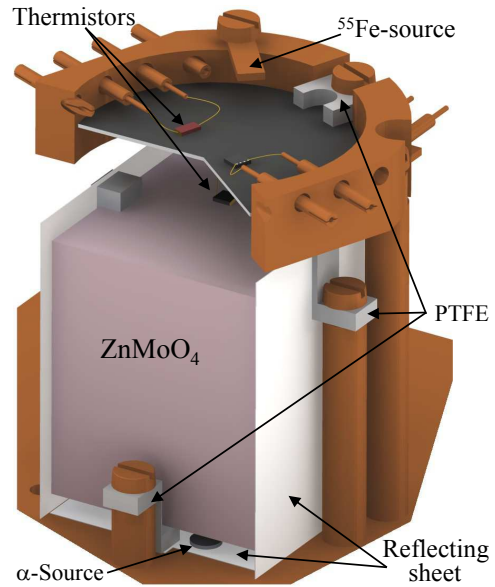}}
\caption{Set-up of the detectors. The ball-bonded Au  wires are crimped into  ``male'' Cu tubes (pins) and inserted into ground-insulated 
``female'' Cu tubes. Custom wires from detectors towards cryostat are not drawn. A section of the light detector and of the reflecting sheet
is not drawn for a better understanding.}
\label{fig:1}      
\end{figure}
The sample was cut from one of the first large-size crystals grown by the LTG Cz. The development of the growth process of ZnMoO$_4$ 
is at the very beginning, and the shape of the crystal is non-optimal. The final form of the crystal was chosen as a compromise between
a large crystal size and the minimization of visible defects.
The shape of the crystal sample used in the current work is an irregular polyhedron with 5
sides whose cross section can be roughly assimilated   to a 45$\times$45 mm$^2$ square.
The only parallel faces are the top and bottom ones.
All the surfaces are matted, except the one faced to the light detector, that was polished.
Unlike previous small samples~\cite{ZMO-Lucifer,Giuliani-2012} that were colorless, this crystal shows an uniform orange tint.

The ZnMoO$_4$ crystal is held by means of four S-shaped PTFE supports fixed to cylindrical Cu columns: two on the top and two on the bottom. 
The crystal is  surrounded laterally and on the bottom part (with no direct thermal contact) by a plastic  reflecting sheet (3M  VM2002).
The temperature sensors are 3x3x1 mm$^3$ Neutron Transmutation Doped (NTD) germanium thermistors, the same used 
in the Cuoricino experiment. For redundancy we decided to use two  thermistors. Each of them is  thermally coupled to the crystal 
via 9 glue spots of $\approx$~0.6~mm diameter and $\approx$~50 $\mu$m height.

At the working temperature of our bolometers (10$\div$30 mK), no  ``standard'' light detectors can work properly.
The best way to overcome this problem is to use a second -very sensitive- ``dark'' bolometer that absorbs the scintillation light giving rise
to a measurable increase of its temperature~\cite{NIMA-2006-A}. 
Our Light Detector (LD) consists of a 50 mm  diameter, 260 $\mu$m  thick pure Ge crystal absorber facing the polished surface of the crystal. 
A schematic view of our set-up is presented in Fig.~\ref{fig:1}.

The detectors were operated deep underground in the Gran Sasso National Laboratories in the CUORE R\&D test cryostat. 
The details of the electronics and the cryogenic facility can be found elsewhere \cite{NIMA-2006-B,NIMA-2006-C,NIMA-2004}.

The heat and light pulses, produced by a particle interacting in  the absorber and transduced in a voltage pulse by 
the NTD thermistors, are amplified and fed into an 18-bits NI-6284 PXI ADC unit.
The trigger is software generated on each thermistor and when it fires  1~s long waveforms, sampled at 2 kHz,  are then saved on disk.
The time window for the LD is shorter and corresponds to 250 ms. 
Moreover, when the trigger of a ZnMoO$_4$ thermistor  fires, the corresponding waveform from the LD is recorded,  irrespective of its trigger.

As one of the main goals of the measurements was  to  test  the $\alpha$ vs $\beta/\gamma$ discrimination capability of this large crystal, a 
$^{238}$U/$^{234}$U $\alpha$ source was faced to the crystals, on the opposite side  with respect to the LD.
The  source  was covered with a 12 $\mu$m thick polyethylene  film, in order  to smear the $\alpha$'s  energies 
down to the $^{100}$Mo Q$_{\beta\beta}$-value.
Since mounted close to  the detector, this source is responsible for an increase of the $\alpha$ background that could in principle spoil our
sensitivity to the intrinsic contamination of the ZnMoO$_4$ crystal. However, the straggling of alpha particles inside the 
polyethylene film ``shifts'' the $^{238}$U/$^{234}$U  $\alpha$-particles toward lower energies (i.e. below 4 MeV)  removing a
possible interference with the 4 MeV peak that should appear in the case of a $^{232}$Th bulk contamination of the crystal.

The $\gamma$ calibration of the  ZnMoO$_4$ crystal is performed through  removable $^{228}$Th and $^{40}$K   sources inserted between the dewar 
housing the cryostat and the external lead shield.
The energy calibration of the LD is achieved thanks to a permanent $^{55}$Fe X-ray source, producing  two X-rays at 5.9 and 
6.5~keV,  faced  closely to the LD.
\subsection{Data Analysis}
\label{Data-analysis}
The amplitude and the shape of the voltage pulse is determined by the off-line analysis that makes use of the Optimum Filter 
technique~\cite{GattiManfredi,Radeka:1966}. The signal amplitudes are computed as the maximum of the filtered pulse.
The amplitude of the light signal is estimated from the value of the filtered waveform at a fixed time delay with respect to 
the signal of the ZnMoO$_4$ bolometer, as described in detail in Ref.~\cite{Piperno:2011fp}.
The signal shape is evaluated on the basis of four different parameters: $\tau_{R}$, $\tau_{D}$, TVL and TVR. 
$\tau_{R}$ (the rise time) and $\tau_{D}$ (the decay time) are evaluated on the raw pulse as (t$_{90\%}$-t$_{10\%}$) 
and (t$_{30\%}$-t$_{90\%}$) respectively. 
TVR (Test Value Right) and TVL (Test Value Left) are computed on the  filtered pulse as the least square differences with respect 
to the filtered response function\footnote{The response function of the detector, i.e. the shape of a pulse in absence of 
noise, is estimated from the average of a large number of raw pulses. It is also used, together with the measured noise power spectrum, to construct 
the transfer function of the Optimum Filter.} of the detector: TVR on the right and TVL on the left side of the optimally filtered pulse maximum. 
These two parameters do not have a direct physical meaning, however they are extremely sensitive (even in noisy conditions) to any difference 
between the shape of the analyzed pulse and the response function.
\begin{table}[htb]
\centering
\caption{Technical details for the ZnMoO$_4$ bolometer (Thermistor 1 and Thermistor 2) and for the LD.
 Signal represents  the absolute voltage drop across the thermistor for a unitary energy deposition.}
\label{Table:parameters_measurement}
\begin{tabular}{lccccc}
\hline
Crystal           &Signal             &FWHM$_{base}$      &$\tau_{R}$    &$\tau_{D}$	 \\
                 &[$\mu$V/MeV]  	       &[keV]              &[ms]          &[ms]   	 \\
\hline
ZnMoO$_4$-1          &  16  			    & 3.6             & 12.8 		      & 59.8 \\
ZnMoO$_4$-2          &  14     	     	    & 3.7             & 12.0  		      & 60.3 \\
LD                   & 1800    				& 0.20            & 3.2 		      & 8.2  \\
\hline
\end{tabular}
\end{table}
\begin{figure}
\resizebox{0.48\textwidth}{!}{%
\includegraphics{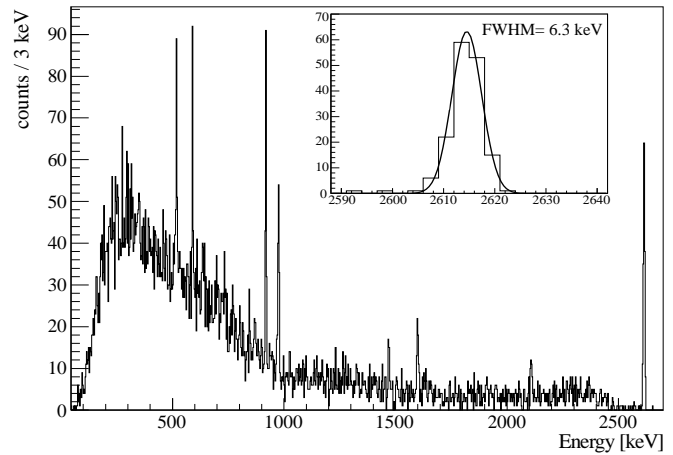}}
\caption{Calibration spectrum  obtained by exposing the ZnMoO$_4$ crystal to the  $^{228}$Th  source for 80 h.
The peak at 2615 keV of $^{208}$Tl, magnified in the inset, shows a FWHM resolution of 6.3 keV.}
\label{fig:2}      
\end{figure}
The detector performances  are reported in Tab.~\ref{Table:parameters_measurement}. 
The baseline resolution, FWHM$_{base}$, is governed by the noise fluctuation at the filter output, and does not depend on the absolute pulse 
amplitude. 

As mentioned above, the use of two thermistors on the same absorber is often made for redundancy. 
In this case we took advantage of the similar performance of both of them, using their sum. 
This technique is useful in the case that the noise fluctuations of the two thermistors are not correlated, meaning that these fluctuations
are not actual temperature fluctuation of the crystal.
This technique can be used in two different ways. One can   off-line  combine the energies measured by the two thermistors into a ``weighted'' 
energy estimator~\cite{cherenkov}, linear combination of the two thermistors.
Or, as in this case, one can  sum the two signals at hardware level and treat the obtained  signal as an additional independent channel.
In our case the sum is performed after the two signals are amplified and just before they  are fed into the acquisition.
The calibration spectrum obtained on the sum of the two thermistors  is presented in Fig.~\ref{fig:2}.

The baseline energy resolution, FWHM$_{base}$ evaluated on the sum (ZnMoO$_4$-Sum) is 2.6 keV, slightly better with respect 
to the ones reported in Tab.~\ref{Table:parameters_measurement}. 
The FWHM energy resolutions  obtained at different energies  are reported in Tab.~\ref{Table:Energy-Resolutions}.

\begin{table}[htb]
\centering
\caption{FWHM energy resolutions of the ZnMoO$_4$ detector evaluated on the two thermistors and on their sum.}
\label{Table:Energy-Resolutions}
\begin{tabular}{lccccc}
\hline
	          &		   ZnMoO$_4$-1	&	ZnMoO$_4$-2		   	   &	ZnMoO$_4$-Sum 		\\
              &		   [keV]		&	 [keV]      	   	   &	    [keV]			\\
\hline
583 keV		  &		  4.1$\pm$0.7 	&	 3.0$\pm$0.5	   	   &	2.9$\pm$0.4 		\\
911 keV		  &		  4.9$\pm$0.4 	&	 4.7$\pm$0.5		   &	4.0$\pm$0.4			\\     
1461 keV	  &		  4.9$\pm$1.5 	&	 5.4$\pm$1.2		   &	4.9$\pm$1.0		    \\
2615 keV	  &		  6.8$\pm$0.4 	&	 6.6$\pm$0.6		   &	6.3$\pm$0.5  		\\
\hline
\end{tabular}
\end{table}

\section{$\alpha$ vs $\beta/\gamma$  discrimination}
\label{sec:discrimination}
As described in Sec.~\ref{sec:introduction}, the possibility to discriminate the $\alpha$ interaction results to be the actual 
key point for a DBD bolometer.
As described in details in~\cite{ZMO-Lucifer}, in scintillating ZnMoO$_4$ bolometers the $\alpha$ vs $\beta/\gamma$  discrimination
can be obtained in two different ways: using the light signal and/or by using the PSD. 
We define  discrimination power (DP) between  the $\alpha$  and $\beta/\gamma$ distributions    the difference between the average values of the two 
distributions  normalized to the square root of the quadratic sum of  their  widths: 
\begin{equation} 
\rm{DP} = \frac{\mu_{\beta/\gamma}-\mu_{\alpha}}{\sqrt{\sigma_{\beta/\gamma}^2+\sigma_{\alpha}^2}}.
\label{eq:DP}
\end{equation} 
\subsection{Light vs heat discrimination}
\label{light-discrimination}
\begin{figure}[t]
\resizebox{0.48\textwidth}{!}{%
\includegraphics{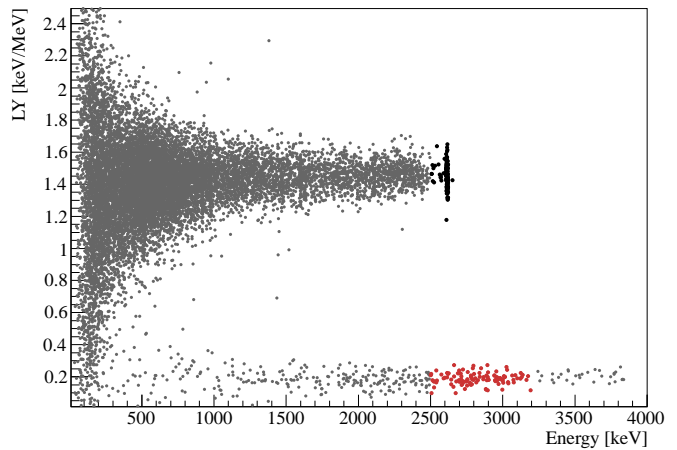}}
\caption{The light-to-heat energy ratio as a function of the heat energy obtained in the 80 h $^{228}$Th calibration with  ZnMoO$_4$-Sum. 
The upper band (ascribed to $\beta/\gamma$ events)  and  lower band (populated by $\alpha$ decays) are clearly separated. The  2615 
keV  $^{208}$Tl $\gamma$-line is well visible in the   $\beta/\gamma$ band as well as a the continuous background induced by the 
degraded $\alpha$ source. The events belonging to the energy region 2.5$\div$3.2 MeV (highlighted in the plot) 
are used to evaluate the DP, that results $\approx$~19.}
\label{fig:3}      
\end{figure}

The light-to-heat energy  ratio~\footnote{Since we attribute to the heat peaks the nominal energy of the 
calibration $\gamma$'s, the  light-to-heat energy  ratio also represents the Light Yield of the crystal.} as a function of 
the heat energy is shown for the calibration spectrum in Fig.~\ref{fig:3}.
$\beta/\gamma$ and $\alpha$ decays give rise to very clear separate distributions. 
In the upper band, ascribed to  $\beta/\gamma$ events, the  2615 keV  $\gamma$-line is well visible. 
The lower band, populated by  $\alpha$ decays, shows  the continuous background induced by the degraded $\alpha$ source.

The evaluated LY of the ZnMoO$_4$ crystal  is 1.54$\pm$0.01 keV/MeV. This value is surprisingly larger (+ 40\%) 
with respect to our previous  measurements  on 30 g (colorless) samples~\cite{ZMO-Lucifer}. This, unexpected, larger LY results in an increase of the 
DP using the scintillation light. Considering the events in the 2.5$\div$3.2 MeV region (see Fig.~\ref{fig:3}) we estimate a DP of  $\approx$~19.
Moreover the scintillation yield evaluated on the internal $\alpha$-line of  $^{210}$Po (see Sec.~\ref{sec:internal-contaminations}) is
0.257$\pm$0.002 keV/MeV that corresponds to a scintillation Quenching Factor of 0.167$\pm$0.002. This value is fully compatible with 
the one (0.18$\pm$0.02) obtained on the 30 g  sample.

\begin{figure}
\centering
\resizebox{0.48\textwidth}{!}{%
\includegraphics{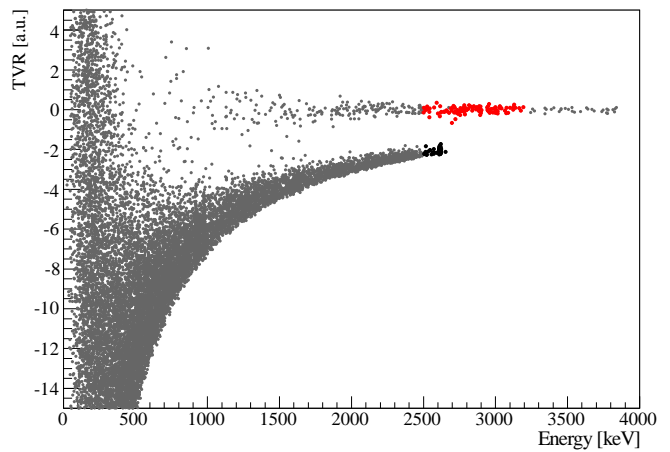}}
\caption{TVR as a function of the energy, for the same events of Fig.~\ref{fig:3}. The upper band is populated by  
$\alpha$ particles (events in the 2.5$\div$3.2 MeV energy range   are shown in red)  while  $\beta/\gamma$'s contribute to the lower band (events in the 
2.5$\div$3.2 MeV energy range  are shown in black).}
\label{fig:4}      
\end{figure}

\begin{figure}
\centering
\resizebox{0.48\textwidth}{!}{%
\includegraphics{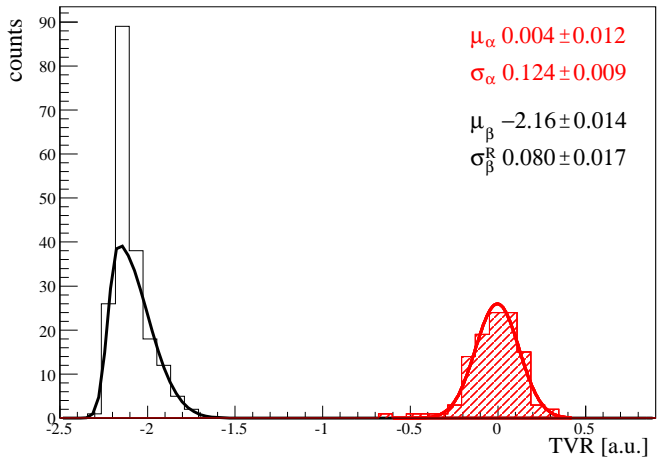}}
\caption{TVR histogram of the events of Fig.~\ref{fig:4}: the $\alpha$ sample in red and  the $\beta/\gamma$ in black. The mean values and  the standard 
deviations, as estimated  from a Gaussian fit, are reported. The $\beta/\gamma$ distribution was fit with an asimmetric Gaussian. 
The sigma of the right part of the gaussian ($\sigma^R_\beta$)is used to evaluate the DP. From Eq.~\ref{eq:DP} a discrimination power 
of $\approx$~14 is obtained.}
\label{fig:5}      
\end{figure}

\subsection{Pulse Shape Discrimination}
\label{PSD-discrimination}
The PSD performed in this work is obtained with the same method described in~\cite{ZMO-Lucifer}.
In Fig.~\ref{fig:4} we plot the   TVR variable as a function of the energy for the same data sample of Fig.\ref{fig:3}.  
As for the case of  Fig.\ref{fig:3}, $\beta/\gamma$ and $\alpha$ events are distinctly separated.
The obtained DP with the  PSD is $\approx$~14, as shown in Fig.~\ref{fig:5}. It has to be pointed out the ``opposite'' behaviour of the discrimination power with respect to the 
results obtained on the 30 g small samples~\cite{ZMO-Lucifer} in which we obtained $\approx$~8 with the light signal and $\approx$~20 with the PSD.
This is due to two distinct mechanisms. First the crystal tested in this  work (despite its larger size and orange tint) 
emits $\approx$~40~\% more light with respect to the
small (colorless) sample previously tested. This implies, obviously,  an improved DP of the scintillation light. Second the 
PSD is sensitive to the S/N ratio. In this work the FWHM$_{base}$ resolution  is 4 times worse with respect to the sample previously tested.

\section{Internal contaminations}
\label{sec:internal-contaminations}
\begin{figure}
\centering
\resizebox{0.48\textwidth}{!}{%
\includegraphics{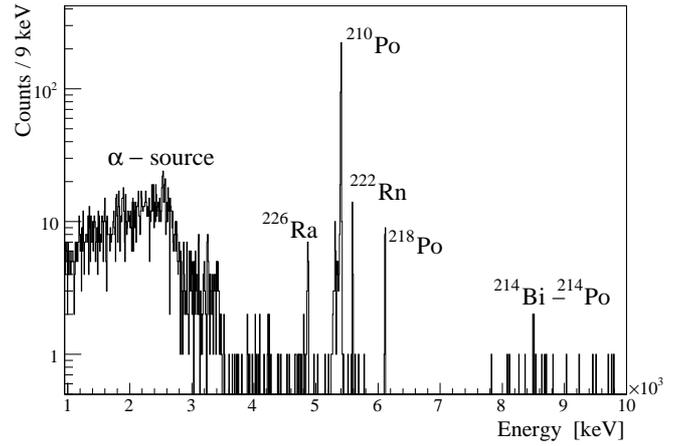}}
\caption{$\alpha$-spectrum obtained in 524 h of measurement. The contribution of the $^{238}$U/$^{234}$U $\alpha$ source  is clearly evident below 4 MeV.
The internal $\alpha$-lines arising from  $^{226}$Ra decay chain are highlighted.}
\label{fig:6}      
\end{figure}

The internal radioactive contaminations of this crystal  were evaluated summing up background and different calibration runs for a total
collected statistics of 524 h. The corresponding $\alpha$-spectrum is presented in Fig.~\ref{fig:6}.
We found a  contamination of  $^{226}$Ra (it  shows a very clear  $\alpha$ and ``BiPo'' decay pattern sequence). This contamination is evaluated as 
27$\pm$6 $\mu$Bq/kg. As often happens, we also found a clear internal contamination of $^{210}$Po, corresponding to an activity of 
700$\pm$30 $\mu$Bq/kg. No other $\alpha$ lines appear in the spectrum. 

In order to evaluate the limits on other potential dangerous nuclei (in particular the $^{232}$Th chain), we  evaluated first the flat 
$\alpha$ continuum in an energy region in which no peaks are expected (3.6$\div$4 and   4.35$\div$4.7 MeV).
Then we studied an interval of $\pm 3\sigma$ centered around the Q value of each possible radioactive nucleus, being $\sigma$ the energy resolution 
of the $^{210}$Po peak (4.5 keV). The expected flat background contribution in each 27 keV energy window is 0.68 counts.
Applying the Feldman-Cousin method~\cite{Feldman-Cousins} using the observed number of counts in each energy window with the expected 
background, we were able to set  90\% CL limits on several nuclei, as reported in Tab.~\ref{Tab:3}. 

\begin{table}[b]
\begin{center}
\begin{tabular}{lcccc}
\hline
Chain & nuclide   & activity\\
      &           & $\mu$Bq/kg \\ 
\hline
$^{232}$Th  & $^{232}$Th  & $<$ 8 	   \\
            & $^{228}$Th  & $<$ 6 	   \\
\hline
$^{238}$U  & $^{238}$U    & $<$ 6  	   \\
           & $^{234}$U    & $<$ 11     \\
           & $^{230}$Th   & $<$ 6 	   \\
           & $^{226}$Ra   & 27$\pm$6   \\
           & $^{210}$Po   & 700$\pm$30 \\
\hline
\end{tabular}
\end{center}
\caption{Evaluated internal radioactive contaminations. Limits are at 90\% CL.} 
\label{Tab:3}
\end{table}

\section{Conclusions}
For the first time a large mass ZnMoO$_4$ crystal was tested as a scintillating bolometer for a possible next generation neutrinoless double 
beta decay experiment. The bolometer shows an excellent energy resolution.
We demonstrated that, even on large mass detector, this compound is able to discriminate $\alpha$ particles interactions 
at -practically- any  desirable level, using the light information as well as the pulse shape discrimination alone.
Moreover this crystal shows an excellent radiopurity.

\section{Acknowledgements}
This work was made in the frame of the LUCIFER experiment, funded by the European Research Council under the European Union's Seventh Framework 
Programme (FP7/2007-2013)/ERC grant agreement n. 247115. 
Thanks are due to the LNGS mechanical workshop and in particular to E. Tatananni, A. Rotilio, A. Corsi, and B. Romualdi   for continuous and 
constructive help in the  overall set-up construction. 
Finally, we are especially grateful to Maurizio Perego for his invaluable help.

\end{document}